\documentclass[12pt,english]{revtex4}
\usepackage[T1]{fontenc}
\usepackage[latin1]{inputenc}
\usepackage{geometry}
\usepackage{amsmath}
\usepackage{babel}
\usepackage{graphics}
\usepackage{amssymb}

\makeatletter

\providecommand{\LyX}{L\kern-.1667em\lower.25em\hbox{Y}\kern-.125emX\@}

\makeatother
\begin{document}

\title{Supersymmetry and Identical Bands}

\author{P.~von Brentano}

\affiliation{Institut f\"{u}r Kernphysik, Universit\"{a}t zu Koeln,
Germany }

\begin{abstract}
Supersymmetry as applied to identical bands is discussed.
A review of the work of the Koeln-Dubna group on this topic
is given and examples in \( ^{171,172} \)Yb , \( ^{173,174} \)Hf
and \( ^{195,194} \)Pt are discussed. The role of pseudo-spin
in the supersymmetry is investigated. A recent precision
lifetime measurement for identical bands in \( ^{171,172} \)Yb
is discussed.  Keywords: U(6/12) supersymmetry, identical
bands, pseudo-spin \\
 \vspace{1cm}
\end{abstract}
\maketitle
\indent

\section{Introduction}

Supersymmetry has been successfully transferred from particle
physics to nuclear physics by Franco Iachello and coworkers\cite{Ia1,Ia2,Frank,Ia3,Balen,BiIa1,U6,Jol1,WaCa1,Ve,VanIs1}.
Various attempts to obtain an extended classification,
in which several neighboring nuclei including even, odd
and odd--odd ones are described by the same Hamiltonian,
have led to convincing results e.g.\cite{JolMe2}. However,
in nuclear structure theory primary attention has been
paid to the investigation of models with Hamiltonians which
exhibit dynamical supersymmetries. In a recent paper\cite{Jol99}
it has been shown, however, that the concepts of identical
bands\cite{Byrski,Nazarewicz,Baktash} and pseudo-spin
\cite{Arima,Hecht} allows one to discuss (partial) supersymmetry
also for cases in which the even partner nucleus has no
dynamical symmetry. This has also been discussed in the
consistent Q formalism\cite{consQ1,consQ2}. The concept
of {}``Identical Bands'' (IB) was suggested by a Strasbourgh-Liverpool
collaboration led by Byrski \cite{Byrski}, who found an
agreement to better than 0.3\% in the gamma energies of
two superdeformed (sd) bands in \( ^{151} \)Tb and\( ^{152} \)Dy.
The role of pseudo-spin in identical bands was subsequently
clarified by Nazarewicz et al. \cite{Nazarewicz}. An explanation
based on the U(6/12) supersymmetry was given in the frame
of the Interacting Boson Model by Gelberg et  al. \cite{Gelberg B}
and in the frame of supersymmetric quantum mechanics by
Amado, Bijker et  al. \cite{Amado}. Although the importance
of the discovery of the IB was widely recognized the data
on the Dy-Tb pair are still incomplete due to experimental
difficulties. The group at Argonne has recently settled
the spins of the identical superdeformed band in \( ^{152} \)Dy
\cite{ANL} but the spins in \( ^{151} \)Tb are still
unknown. Also the 3/2, 7/2... signature partner band in
\( ^{151} \)Tb is still missing. This shows that one has
to rely on the spectroscopy of {}``Identical Bands''
at normal deformation, where spins and parities are all
known. In this case the moments of inertia of the even
and odd nucleus system tend to differ by 10\% or more as
compared to 0.3\% in the SD case. A survey is given by
Baktash et al.\cite{Baktash}. We show in Fig. 1 a convincing
example of the identical bands in \( ^{174} \)Hf and \( ^{173} \)Hf
\cite{Jol99}.

\begin{figure}
{\centering \resizebox*{0.8\columnwidth}{!}{\rotatebox{-90}{\includegraphics{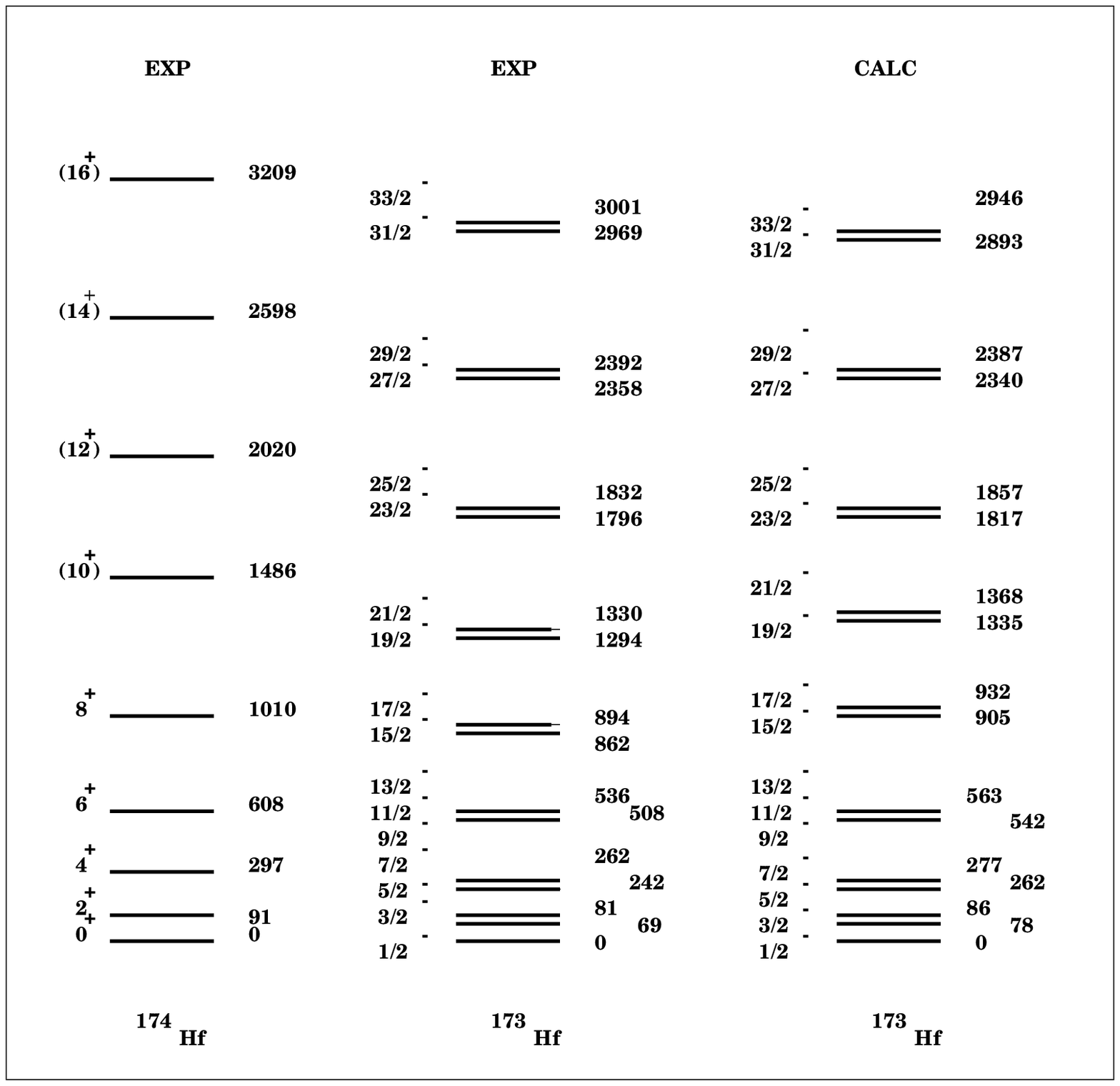}}} \par}

Fig.1

Comparison of the energies of SUSY partner bands in \( ^{174} \)Hf
and\( ^{173} \)Hf. The energies labeled CALC are obtained
by correcting for the change in the moment of inertia and
by allowing a pseudo-spin orbit coupling. From Jolos et
al.\cite{Jol99}.
\end{figure}

In the following I will review the Koeln-Dubna work\cite{Jol96,Jol99,Jol01,Klug}
which focused on the supersymmetry connected with identical
bands. In order to avoid equations or technical discussions
a simple schematic model for supersymmetry will be discussed.
This is done on the basis of the U(6/12) supersymmetry
\cite{Ia1,U6}which has a very transparent microscopic
foundation. In U(6/12) respectively in U(15/30)\cite{Ia1, U6}
there are three (five) contributing shells in the odd nucleus
and which have the spins 1/2, 3/2 and 5/2 (or 1/2, 3/2,
5/2, 7/2 and 9/2)

\section{Pseudo-spin}

In discussing the U(6/12) SUSY and the schematic model
pseudo-spin plays an important role. Thus I make a few
comments on pseudo-spin symmetry. This approximate symmetry
is long known \cite{Arima,Hecht}. Recently it was shown,
that pseudo-spin symmetry is actually based on a relativistic
symmetry\cite{GinoPS}. We remind briefly the pseudo-spin
scheme: In Fig. 2 the s.p. neutron states around \( ^{208} \)Pb
are shown. In the pseudo-spin scheme the 3p1/2, 2f5/2 and
3p3/2 levels are relabeled with the pseudo-spin quantum
numbers 2 \( \widetilde{s} \) 1/2 \textbf{,} 2 \( \widetilde{d} \)
5/2 , 2 \( \widetilde{d} \) 3/2 . Whereas the angular
momenta of the particle are l = 1 or l = 3 respectively,
the pseudo angular momenta are \( \widetilde{l}=0 \) or
\( \widetilde{l}=2 \) . For simplicity it is assumed in
the following that the pseudo-spin of the last odd particle
is fully decoupled in energy. In this case to each state
with spin \( I_{e} \) of the even nucleus there belongs
a multiplet of states in the odd partner nucleus with the
same excitation energy and with spins :

\[
I^{+}_{odd}=I_{e}+1/2\; \&\; \; I^{-}_{odd}=|\; I_{e}-1/2\; |\]

\noindent respectively. In particular there is a doublet
of states for \( I_{e}\neq 0 \) and a singlet for \( I_{e}=0 \).
An example of nearly decoupled pseudo-spin multiplets is
found in very nice data by Jolie and Graw\cite{195Pt}.
They measured a set of states which is presumably complete
below 1 MeV. After this they were able to assign the observed
levels to pseudo-spin multiplets. Indeed - as shown in
Fig. 3 - each level in \( ^{195} \)Pt ( below 0.8 MeV)
comes either as a doublet or as a singlet if \(  \) \(  \)\( I_{e} \)
= 1/2. This is to -my knowledge -
\begin{figure}
{\centering \resizebox*{0.75\columnwidth}{!}{\includegraphics{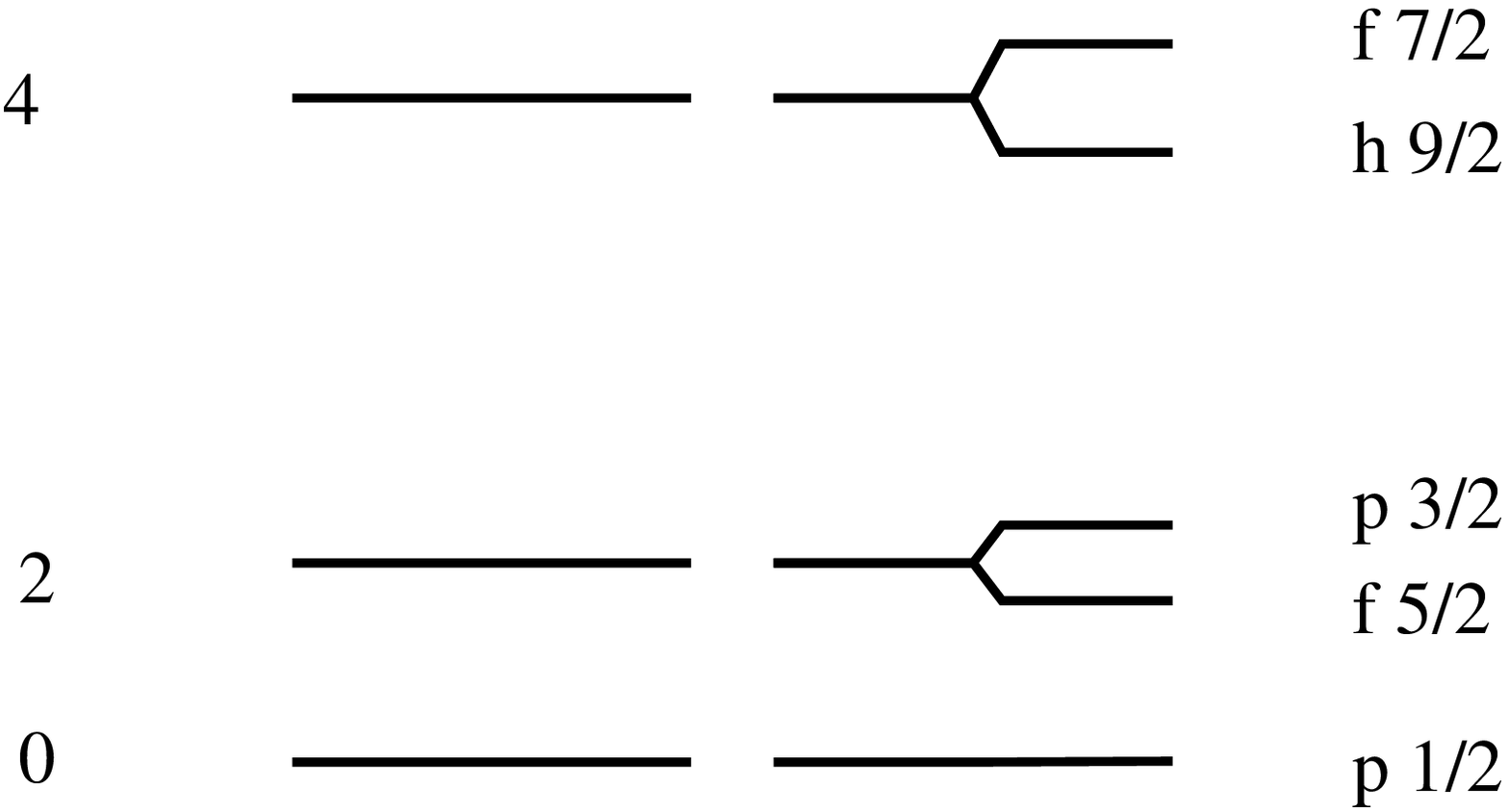}} \par}

Fig. 2

The figure shows on the right side the spins of the neutron
single particle energies in Pb and on the left side their
pseudo angular momentum: \( \widetilde{l} \).
\end{figure}
 the only case in which pseudo-spin multiplets in the odd
nucleus are seen so clearly. So more data are needed. It
is a beautiful prediction from the dynamical U(6/12) supersymmetry
and from the pseudo-spin concept.

\section{Supersymmetry with identical bands : A schematic model.}

I will now discuss a schematic model for supersymmetry
(SUSY) with identical bands in the frame of U(6/12). The
odd particle is in shells with spins 1/2, 3/2 and 5/2 only.
The pseudo angular momenta \( \widetilde{l} \) of the
odd particle have integer values and can be considered
as bosonic excitations. They have the values \( \widetilde{l} \)
= 0 and \( \widetilde{l} \) = 2 corresponding to the spins
L = 0 and L = 2 of the s and d bosons. Thus it is {}``natural''
following T. Otsuka \cite{Otsuka} to describe the particles
with pseudo angular momenta \( \widetilde{l} \) = 0 and
\( \widetilde{l} \) =2 by ( pseudo angular momentum )
quasi bosons with the spins \( \widetilde{l} \) =0 : \( s^{f} \)
and \( \widetilde{l} \)=2 : \( d^{f} \). Thus there are
three kinds of bosons : neutron bosons \( b^{n} \) , proton
bosons \( b^{p} \), and quasi bosons \( b^{f} \) 

The schematic model assumes :

\begin{itemize}
\item Equal energies \( \varepsilon ^{n}_{d}=\varepsilon ^{p}_{d}=\varepsilon _{d}^{f} \)
of the three kinds of bosons : \( b^{n} \) , \( b^{p} \),
\( b^{f} \) 
\item Equal interactions \( V_{ik} \) among the three kinds
of bosons \( b^{n} \) , \( b^{p} \), \( b^{f} \) .
\item Fully decoupled pseudo-spin.
\end{itemize}
These assumptions imply identical bands in the following
three bosonic systems : \( K_{0} \), \( K_{1} \) and
\( K_{2} \) defined in Fig. 3 and which have the same
total boson numbers\( \: \:  \) \( N_{B} \) : but with
different numbers of bosons of the three kinds : \( b^{n} \)
, \( b^{p} \), \( b^{f} \).

\begin{figure}
{\centering \resizebox*{0.85\textwidth}{!}{\includegraphics{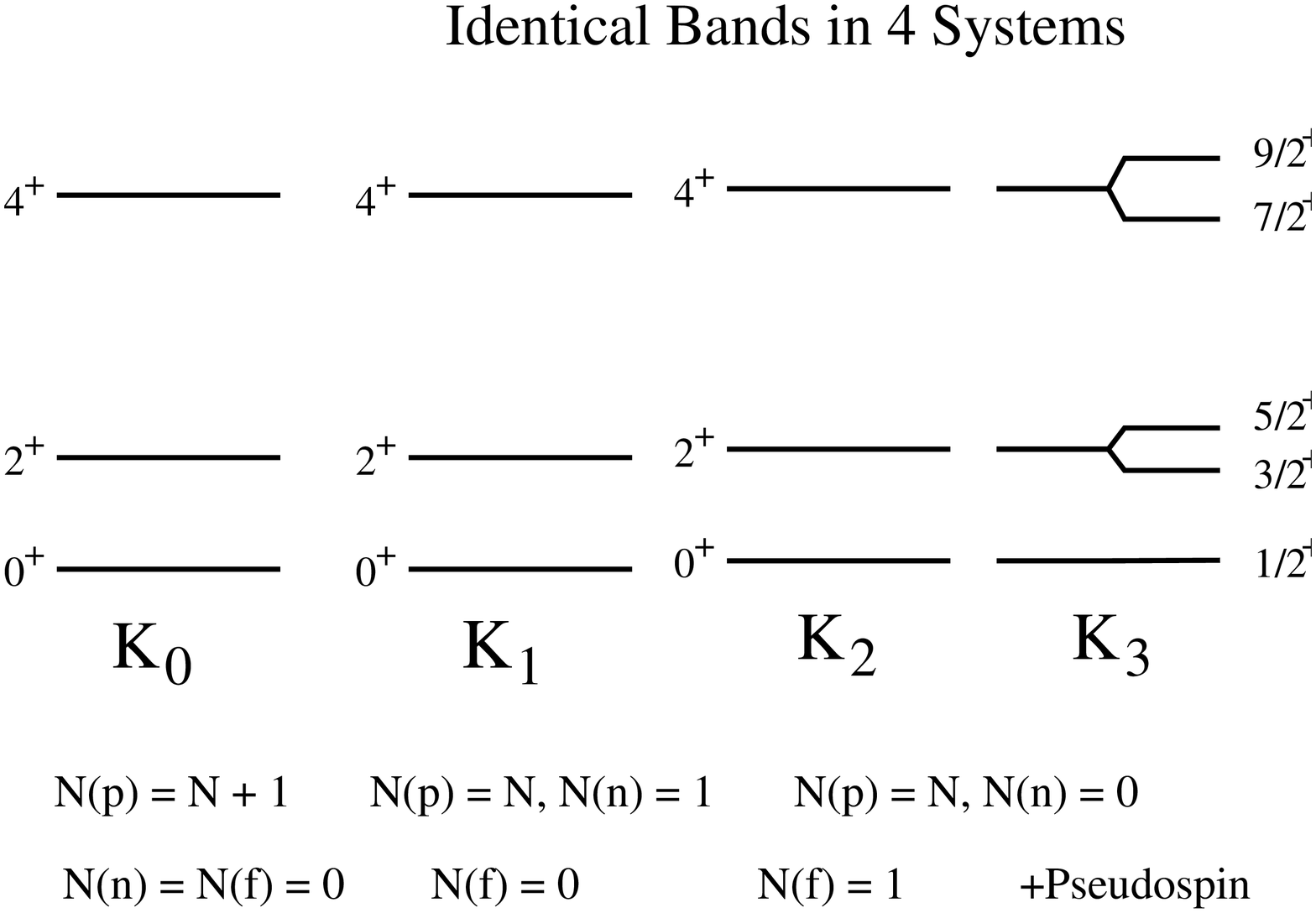}} \par}

Fig. 3

The figure illustrates the {}``Identical Bands'' suggested
by the schematic model in four {}``nuclei'' which have
the same total number of {}``bosons'' but differing numbers
of bosons of various types as shown in the figure. The
{}``nucleus'' \( K_{3} \) is obtained from \( K_{2} \)
by adding the pseudo-spin \( \widetilde{s} \)
\end{figure}

In particular one finds :

\begin{itemize}
\item to every state in \( K_{0} \) there corresponds a state
in \( K_{1} \) with the same energy. \( K_{0} \) and
\( K_{1} \) are members of an F spin multiplet. They have
good F-spin.
\item to every state in \( K_{1} \) there corresponds a state
in \( K_{2} \) with the same energy and vice versa
\item to every state in \( K_{2} \) there corresponds a pseudo-spin
multiplet in \( K_{3} \) with the same energy
\end{itemize}
In summary in the schematic model to each pseudo-spin multiplet
of the odd system there belongs a unique state of the corresponding
even system with the same energy. This is a somewhat surprising
result. It means that there are as many states in the even
{}``nucleus'' \( K_{1} \) as there are pseudo-spin multiplets
in the {}``odd nucleus'' \( K_{3} \). One might think
that the odd nucleus has many more states. One must remember,
however, that also the {}``mixed symmetry'' states in
the {}``even nucleus'' have to be considered.

\section{Comparison of schematic model to data}

Below the following predictions A)... D) of the schematic
model are compared to data:

\begin{itemize}
\item A) to each level below about 0.8 MeV in \( ^{194} \)Pt
there is a multiplet in \( ^{195} \)Pt with similar energy.
We note that not only the yrast band but e.g. also the
gamma band has a SUSY partner. (see Fig. 4 \cite{195Pt})
We note that the energies of the centroids of the doublets
differ somewhat from the energy of the corresponding level
in \( ^{194} \)Pt. Also there is some splitting in the
doublets. Thus the schematic model is broken somewhat
\item B) There are low lying multiplets in \( ^{195} \)Pt which
do not correspond to low lying states in\( ^{194} \)Pt,
however. In the schematic model they might correspond to
high lying mixed symmetry states \cite{BrGe}. An example
is the (199 keV and 224 keV) doublet, which has a core
state with I\( ^{\pi } \) = 1\( ^{+} \).
\end{itemize}

\begin{figure}
{\centering \resizebox*{!}{0.7\textheight}{\includegraphics{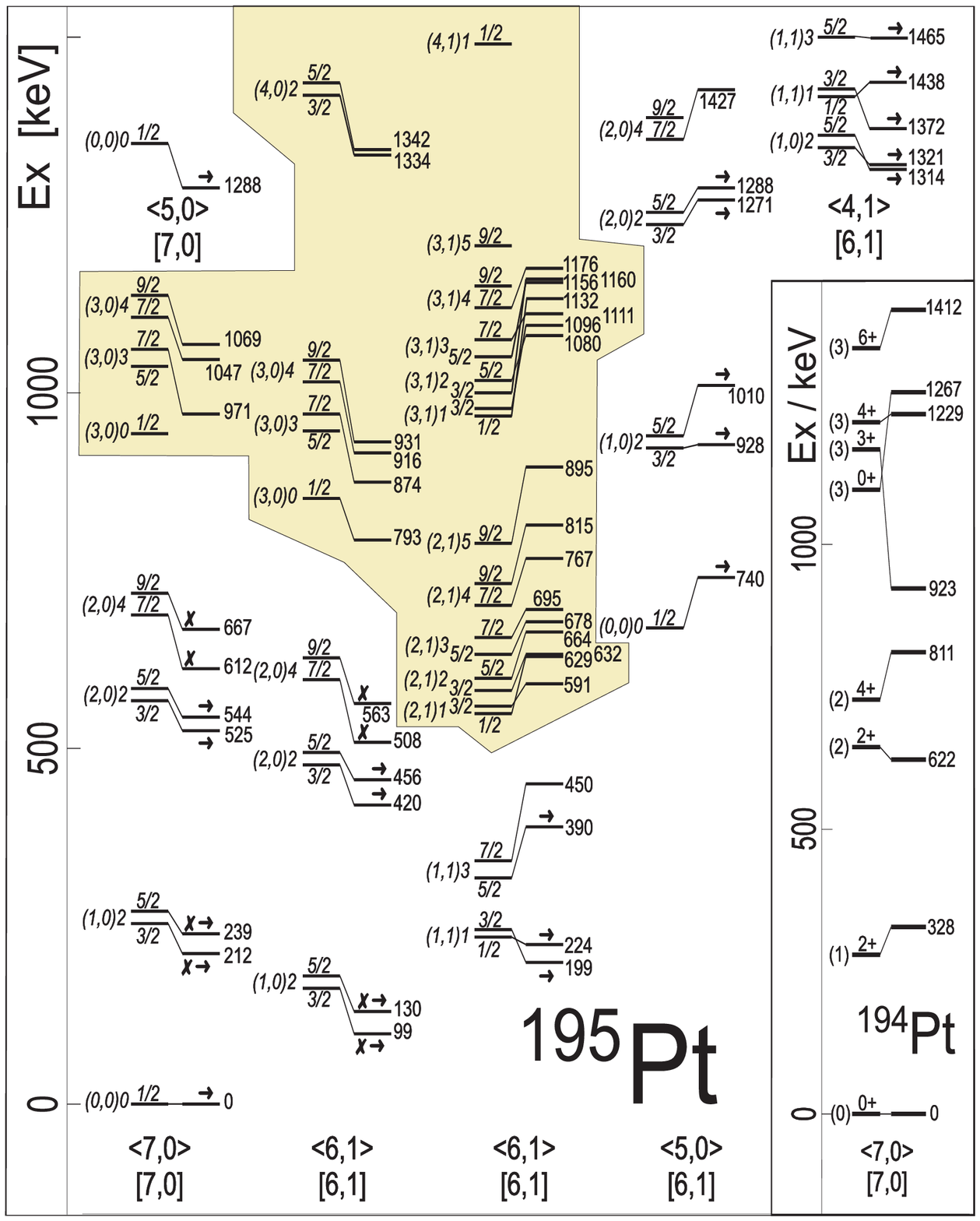}} \par}

Fig. 4

Shown are the spins,experimental(right) and SUSY (left)
energies of \( ^{194} \)Pt and \( ^{195} \)Pt. Each level
below about 0.8 MeV in \( ^{195} \)Pt is a nearly decoupled
pseudo-spin doublet or if \( I=1/2 \) a singlet. The IB
are the states in \( ^{194} \)Pt and the states in \( ^{195} \)Pt
with label \( [7,0] \) below. From Metz et al.\cite{195Pt}
\end{figure}

\begin{itemize}
\item C) in the rare earth region there are identical bands with
a vanishing projection \( \widetilde{L_{3}}=\widetilde{\Lambda }=0 \)
of the pseudo angular momentum on the symmetry axis. The
\( \widetilde{\Lambda }=0 \) band qualifies as super symmetrical
partner band of the K = 0 ground band of the even partner
nucleus. These were the bands considered in \cite{Jol99,Jol01,Klug}
. Examples are identical bands in\( ^{171,172} \)Yb (Fig.
5) , \( ^{173,174} \)Hf ( Fig. 1 ) and in \( ^{181,2} \)Pt
\cite{Jol01}. These are convincing SUSY partner bands.
\item D ) the bands in \( ^{171,172} \)Yb , in \( ^{173,174} \)Hf
and in \( ^{181,1822} \)Pt which have a non vanishing
pseudo angular projection \( \widetilde{L_{3}}=\widetilde{\Lambda }\neq 0 \)
have a pseudo-spin partner band which is separated in energy
by a few hundred keV or more. Thus pseudo-spin is not decoupled
for these bands. And the schematic model breaks down for
these bands .
\end{itemize}
We note that A) and C) show that there are identical bands
and spectra. B) indicates that there are levels in these
nuclei which have no partners with the same energies. Thus
the schematic SUSY survives only for a part of the levels
as a partial symmetry\cite{Alhassid,Leviatan1}. We note,
however, that these extra states are reproduced in the
dynamical U(6/12) SUSY as this uses a different ( more
correct) interaction. Part of the reason for the failure
is certainly that pseudo-spin is not fully decoupled i.e.
the excitation energies in the pseudo-spin doublets are
not equal. In \( ^{207} \)Pb one finds : \( \Delta E \)
(2f5/2 - 3p3/2) = 329keV. However the pseudo-spin splitting
is still much smaller than the spin orbit splitting : \( \Delta E \)
(2f5/2 - 3p3/2) \( \leq  \) 0.36 {*} \( \Delta E \) (3p1/2
-3 p3/2)). An interesting point concerns the apparent validity
of pseudo-spin in the \( \widetilde{\Lambda }=0 \) bands
in \( ^{171,172} \)Yb , \( ^{173,174} \)Hf and \( ^{181,182} \)Pt
(see C) and the fact that pseudo-spin is not decoupled
for the \( \widetilde{\Lambda }\neq 0 \) .The reason may
be that the \( \widetilde{\Lambda }=0 \) state has no
pseudo-spin partner, as is shown e.g. in Fig. 2 for the
3 p 1/2 state, whereas for \( \widetilde{\Lambda }\neq 0 \)
there are two pseudo-spin partner levels e.g. the 2f5/2
and 3p3/2 levels , which can interact. The small decoupling
in the \( \widetilde{\Lambda }=0 \)  bands is also interesting
and deserves a closer study.

\section{Supersymmetry of identical bands in \protect\( ^{171,172}\protect \)Yb
supported by lifetime data}

There are 5 requirements for an K = 0 and an odd rotational
band in neighbor nuclei to be super partners of an unbroken
partial supersymmetry:

1) identity of the moments of inertia,

2a) decoupling of the pseudo-spin.(close doublets in the
odd nucleus),

2b) alternatively a special value a=1 of the decoupling
parameter in the rotational energy formula of the \( K=1/2 \)
band,

3) an even pseudo angular momentum of the particle,

4) identity of the quadrupole transition moments, and

5) \( K=1/2 \), \( \tilde{\Lambda } \)=0 in the odd band.

Some people like to remember these 5 conditions and these
special values of special parameters. Others like to say
there is a SUSY, that implies all.

At this point it might be appropriate to avoid a possible
misunderstanding, namely that the \( K=1/2 \), \( \tilde{\Lambda } \)=0
bands e.g, in \( ^{171} \)Yb could be obtained by a weak
coupling of the spin of the core state I of \( ^{172} \)Yb
to the pseudo-spin.\[
J=\textrm{I}\; \oplus \; \widetilde{s}\]

This is not true at all. Rather one couples the spins I
of several core states strongly to the pseudo angular momentum
\( \widetilde{l} \) of the particle and after this the
pseudo-spin is weakly coupled.\[
J=\Sigma (I,\widetilde{l})\; [\; (\; I\; \oplus \; \widetilde{l\; })\oplus \; \widetilde{s}\; ]\; \]

Thus neither \( I \) nor \( \widetilde{l} \) are not
even approximately good quantum numbers. And although the
\( ^{171,172} \)Yb pair looks like a weak coupling situation,
it is actually a quite complicated coupling scheme: strong
for \( \widetilde{l} \) and weak for \( \widetilde{s} \)
\cite{Jol01}.

Thus it is important to test the SUSY in more detail by
establishing the identity of the quadrupole transitions.
The pair \( ^{171,172} \)Yb is the only case in which
one do Coulomb Excitation on both 
\begin{figure}
{\centering \resizebox*{0.9\textwidth}{!}{\includegraphics{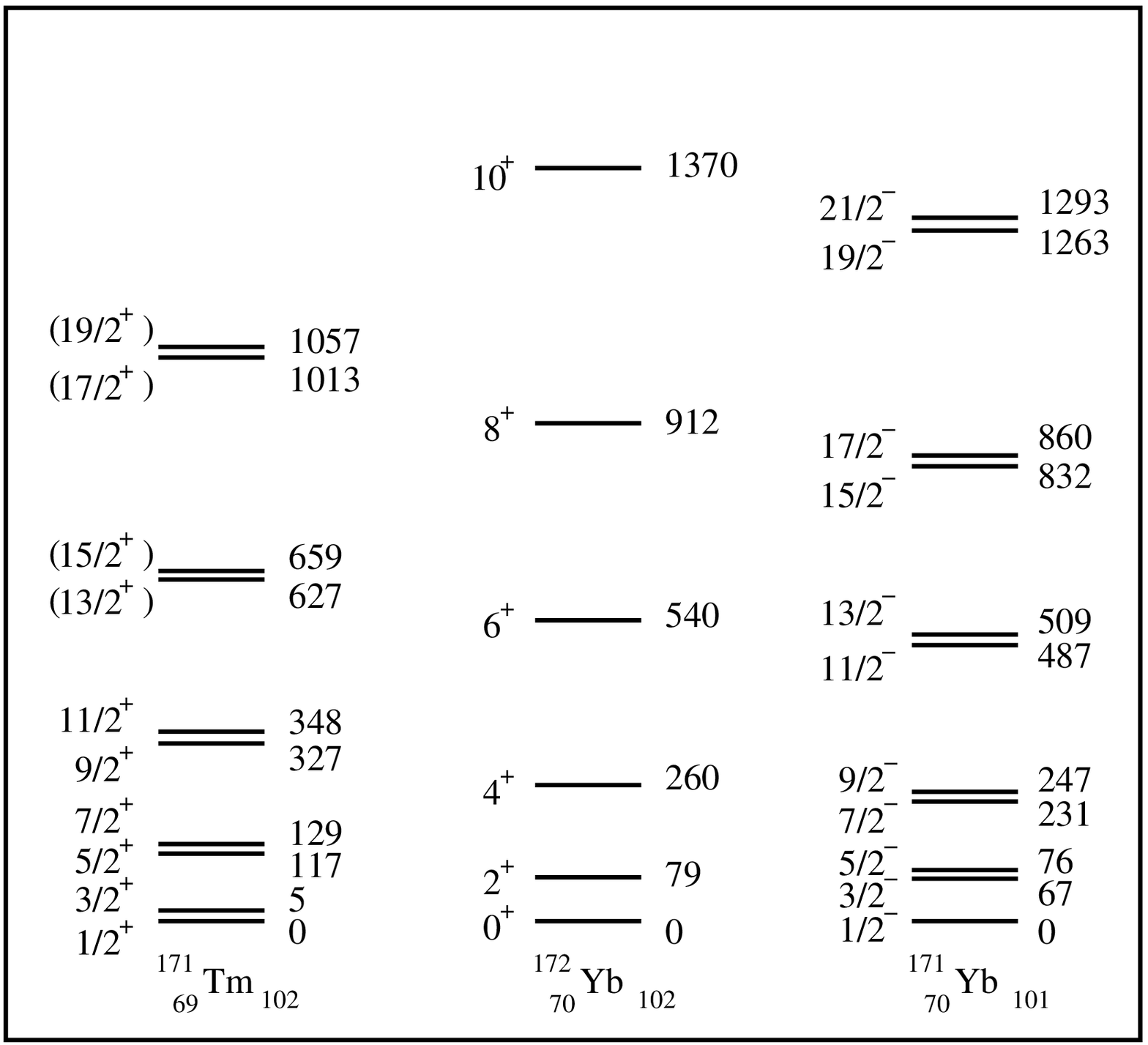}} \par}

Fig. 5

The figure compares K= 1/2 bands in \( ^{171} \)Yb and
\( ^{171} \)Tm with the K = 0 ground band  in \( ^{172} \)Yb
.The bands in Yb have \( \widetilde{\Lambda }=0 \) and
thus form a SUSY doublet . The band in \( ^{171} \)Tm
has \( \widetilde{\Lambda }=1 \) ( doublet for groundstate)
and is thus not the SUSY partner of \( ^{172} \)Yb, which
has \( \widetilde{\Lambda }=0 \). From Klug et al.\cite{Klug}
\end{figure}
for both SUSY partners. The energies are shown in Fig.
5 agree with the SUSY. For these reasons Klug, Dewald et
al. measured precision lifetimes by the recoil distance
method following Coulomb excitation experiments using an
\textsc{Euroball Cluster} detector at the FN-Tandem accelerator
in Koeln \cite{Klug}. From these data and previous data
on branching and mixing ratios from Canberra, the ratios
of the transition quadrupole moments for corresponding
E2 transitions in \( ^{171} \)Yb and \( ^{172} \)Yb were
obtained , which are given in the Table 1.

\begin{table}
Tab.1

Ratios of transition quadrupole moments \( Q_{t} \) from
SUSY partner bands in \( ^{171} \)Yb and\( ^{172} \)Yb.
The \( Q_{t} \) values are obtained from the measured
B(E2) values by using the formulas for the quadrupole moments
for the rigid rotor. From Klug, Dewald et al.\cite{Klug} 

\begin{tabular}{|c|cc|c|c|}
\hline 
&
\( \Delta  \)I=2&
&
\( \Delta I=2 \)&
\( \Delta I=1 \)\\
\hline
I&
\( \frac{Q_{t}^{171}(I+1/2)}{Q_{t}^{172}(I)} \)&
&
\( \frac{Q_{t}^{171}(I-1/2)}{Q_{t}^{172}(I)} \)&
\( \frac{Q_{t}^{171}(I-1/2)}{Q_{t}^{172}(I)} \)\\
\hline 
8&
1.02(6)&
&
1.03(6)&
1.21(10)\\
\hline 
6&
0.94(5)&
&
0.94(5)&
0.95(6)\\
\hline 
4&
0.96(2)&
&
0.95(3)&
0.97(2)\\
\hline
\end{tabular}
\end{table}
One notes that the values of the transition quadrupole
moment \( Q_{t} \) for \( ^{171} \)Yb and \( ^{172} \)Yb
are in equal within an error of a few percent\cite{Klug}.
This is one of the most accurate experimental comparisons
of transition quadrupole moments for supersymmetry partner
bands. This striking agreement supports strongly the proposed
supersymmetry for \( ^{171} \)Yb and \( ^{172} \)Yb.
Indeed all 5 of the requirements for a supersymmetry are
fulfilled for the\( ^{171} \)Yb and \( ^{172} \)Yb pair.
On the contrary in the \( ^{171} \)Tm and \( ^{172} \)Yb
pair only three conditions are fulfilled, whereas the conditions
3) and 5): an even pseudo angular momentum of the particle
are violated. And indeed there is no supersymmetry in this
case.

Summing up convincing examples of (partial) SUSY for energies
and B(E2) values in identical bands are shown. An intuitive
understanding of the U(6/12) SUSY in the frame of the pseudo-spin
concept and the schematic model have been given. Limitations
of supersymmetry and pseudo-spin symmetry restricting the
SUSY to a partial SUSY have been discussed.

\acknowledgments My particular thanks go to Franco Iachello,
for his constant inspiration to Koeln work since many years.
Thanks go also to A. Dewald, A. Gelberg , J. Jolie, R.
V. Jolos and T. Otsuka for joint works published and in
progress on supersymmetry which were used in this talk.
Thanks for many discussions to R. F.  Casten, F. Iachello,
P. van Isacker and N. Pietralla. This work was partially
supported by the DFG under the contract Br 799/10-2.

\end{document}